\documentclass[twocolumn,prl,aps,floats,epsf,psfig,colordvi,showpacs]{revtex4}
\usepackage{array, graphicx}
\newcommand{\numu}{\nu_{\mu}}
\newcommand{\nue}{\nu_{e}}
\newcommand{\nuebar}{\overline{\nu_{e}}}

\newcommand{\numubar}{\overline{\nu}_{\mu}}

\begin{document}
\title{
Muon internal bremsstrahlung: a conventional explanation for the excess
$\nue $ events in MiniBoone
}
\author{ Arie Bodek}
\affiliation{Department of Physics and Astronomy, University of
Rochester, Rochester, NY  14627-0171}
\date{\today}
\begin{abstract}
We show that the rate of charged-current $\numu$ events with a
hard 
internal bremsstrahlung photon is consistent with the excess 
$\nue$  candidate events  reported by the MiniBoone and LSND (decay in
flight) experiments. Hard photons radiated by the muon leg in charged-current  neutrino
interactions ($\numu$N$\rightarrow$$\mu$$\gamma$N)
are a significant source of
background that should be considered by  current and future $\numu$$\rightarrow$$\nue$
neutrino oscillations appearance experiments  (e.g. LSND, MiniBoone, SuperK, MINOS, T2K and NOVA).
\end{abstract}
\maketitle
Experimental evidence for oscillations among the three
neutrino generations was reported almost a decade ago.
The LSND Collaboration\cite{LSND} has also reported evidence for
$\numu$$\rightarrow$$\nue$ oscillations in a  $\Delta~m^{2}$ and mixing
angle region which is not consistent with the atmospheric and
solar neutrino oscillation results.   This LSND result was
obtained from
the observation of  $\nuebar$ events in 
a  $\numubar$ beam  (with energies
between 20 and 55 MeV)  that originates
from decays of muons at rest\cite{LSND}.

The LSND collaboration also searched
for  $\nue$  candidates in an independent
exposure to a  beam of   $\numu$'s
(with energies in the range 60 to 200 MeV)
from pion decay in flight\cite{LSNDdecay}.
The observed rate  of
$\nue$ events in the decay in  flight
sample  has been reported to be consistent with the oscillations
parameters from the decay at rest 
results (though with a lower statistical significance).

The MiniBoone experiment was constructed to investigate
the origin of the anomalous excess of 
$\nue$  events reported by LSND. The 
$\numu$ and $\numubar$ beams at MiniBoone are
at higher energy  (0.2-3.0 GeV)

The first published results\cite{boone} from MiniBoone
(taken with a  $\numu$ beam) do not confirm the LSND results. 
 The observed yield and energy spectrum of 
the MiniBoone  $\nue$ candidates (shown in figure 1)  is incompatible
with  the oscillation parameters in the LSND allowed region.
However, the MiniBoone data show an excess of $\nue$ candidates 
at energies below their analysis  threshold of $475~MeV$.
The total excess  which is not understood, is about 150 events in a sample
of $\approx$ 400,000 $\numu$ charged current events, or about 0.03$\%$.
When the analysis is extended to lower energy \cite{booneLP} (see 
in figure 2), the excess of $\nue$ events persists.

In this communication we show that the calculated rate of 
charged-current $\numu$ neutrino events with a
hard  internal bremsstrahlung photon is consistent with the excess 
$\nue$ event candidates reported by MiniBoone. 
Hard photons radiated by the muon leg in charged 
current $\numu$  events ( $\numu$N$\rightarrow$$\mu$$\gamma$N)
are a significant source of
background that should be considered by $\numu$$\rightarrow$$\nue$
neutrino oscillations appearance experiments.

\begin{figure}
 \begin{center}
\includegraphics[width=3.5in,
height=3.8in]{{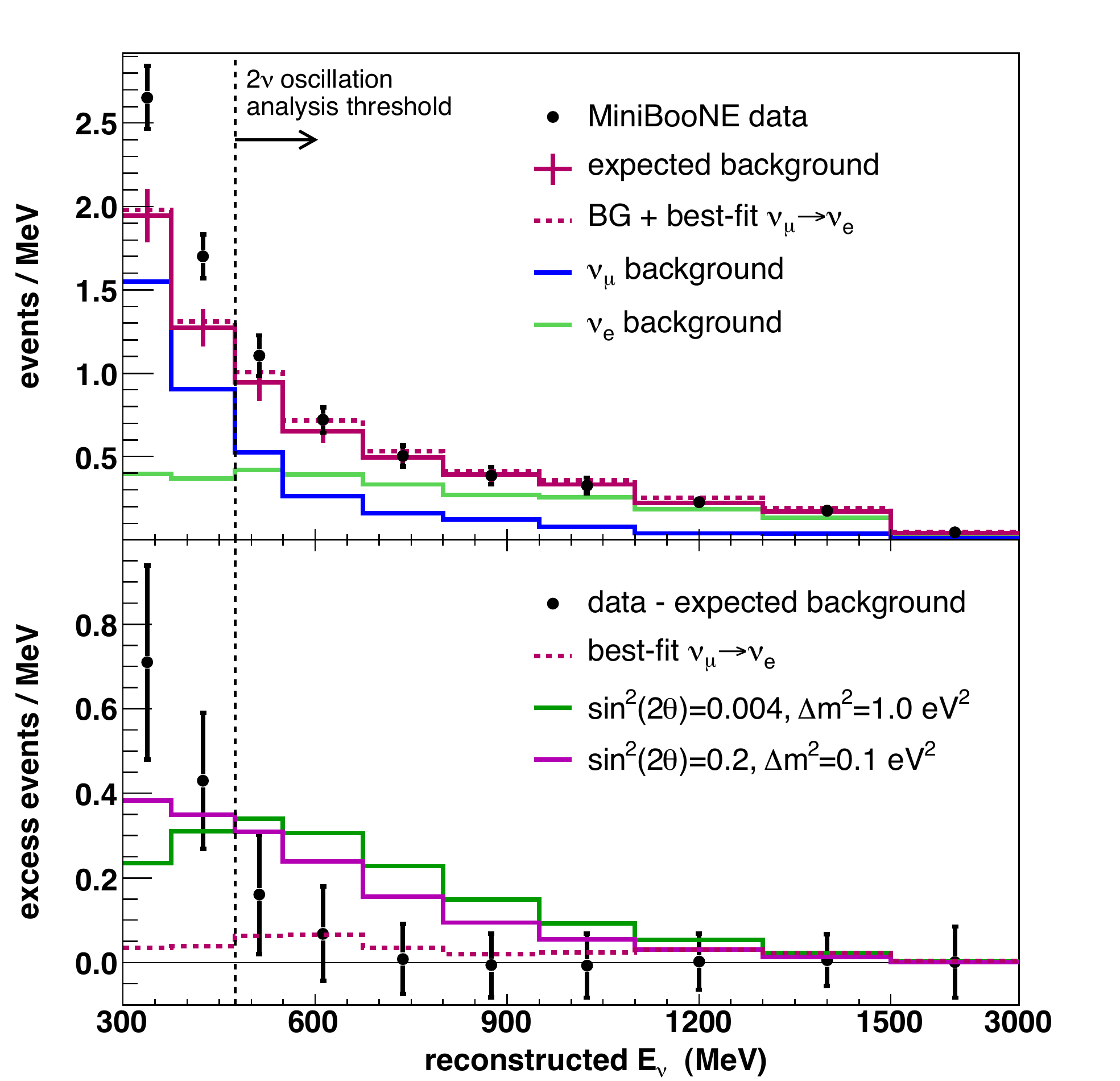}}
\caption[energy]{ The  reconstructed neutrino energy ($E_{\nu}^ {rec}$)
 distribution of the total (top) and  excess (bottom)  of $\nue$
charged current quasi-elastic candidate events published by MiniBoone. 
 The total excess, which is not understood (below their analysis  threshold of $475~MeV$) is about 150 events  in a sample of $\approx$ 400,000 $\numu$  charged current events, or about 0.03$\%$ .
Also shown are the best fit oscillation parameters, and the predicted excess
from  oscillation parameters in the LSND allowed region.
}
 \end{center}
     \label{boonefig}  
\end{figure}

The probability for radiation of a hard photon by muons created in a quasielastic
$\numu$  charged current interaction in the MiniBoone energy range is about  $1\%$. 
The probability to radiate a photon which carries about half of the muon energy
is about $0.3\%$.  Such high energy photons appear as electromagnetic (EM)  showers in
Cerenkov detectors such as MiniBoone.  In Cerenkov detectors, muons with energy
less than 200 MeV are practically invisible, since they are below Cerenkov threhold.
Therefore, 400 MeV muons, which radiates a 200 MeV photon would be invisible
in MiniBoone and appear as  $\nue$ candidates.
Even if a muon is partially visible in MiniBoone because its
energy is above Cerenkov threshold, or due to the production of a
low light level from the scintillator in oil (about $25\%$),
its signal  would most likely be buried within the  photon electromagnetic shower.
The presence of some of these undetected muons may be inferred in MiniBoone  
via the  observation of a Michel ($<$ 52.8  MeV) electron from the delayed decay  at
rest  of  the muon ($\mu-DAR$).  However, the efficiency for the detection of
the presence of such muon events using  $\mu-DAR$ photons  is not $100\%$
for several reasons: 

(1) A fraction of of the DAR electrons are
at very low electron energy. 

(2)  A fraction  of the muons decay outside the detector's active fiducial volume.

(3) A fraction of the  negatively charged muons are captured by atomic nuclei and undergo
internal conversion. 

In this communication, we show that the rate of radiative muon events
is significant and should be included in the MiniBoone analysis. 
The details of the inefficiency  in the detection of invisible or partially visible
muons depend on the  specific details of the  MiniBoone analysis,
and are not addressed in this communication. 

\begin{figure}
 \begin{center}
\includegraphics[width=3.5in,
height=3.8in]{{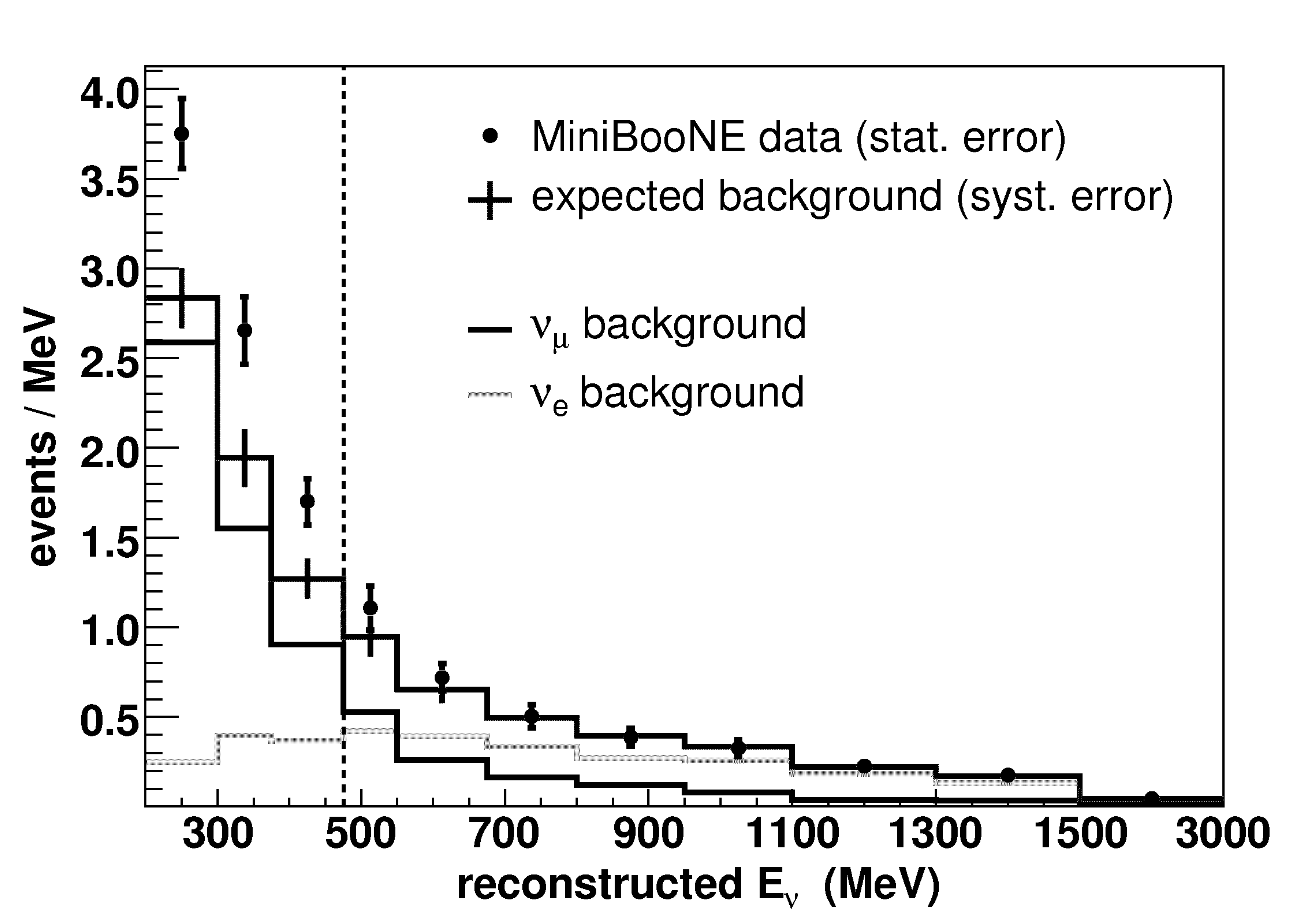}}
\caption[energy]{The  reconstructed neutrino energy ($E_{\nu}^ {rec}$)
 distribution of  excess  of $\nue$ from an updated
 analysis extended to lower energy.
}
 \end{center}
     \label{boonefig2}  
\end{figure}

The radiative corrections to inclusive charged current neutrino scattering
has been  calculated by by several authors\cite{radcor}  including Kiskis,
Barlow and Wolfram, and
 Bardin $et~al$. In this communication, we use the leading
 log peaking approximation as  derived by  De R\'{u}jula, Petronzio, and Savoy-Navarro \cite{rujula}.

\begin{figure}
 \begin{center}
\includegraphics[width=1.5in]{{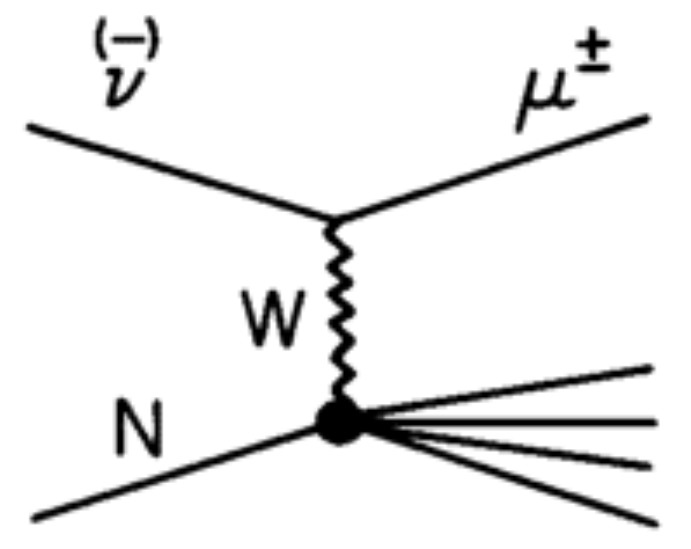}}
\caption[energy]{  The the lowest order Born diagram for $d^2\sigma_0/dxdy$ for
$\numu$N$\rightarrow \mu$N}
 \end{center}
     \label{fig3}  
\end{figure}
\begin{figure}
 \begin{center}
\includegraphics[width=1.5in]{{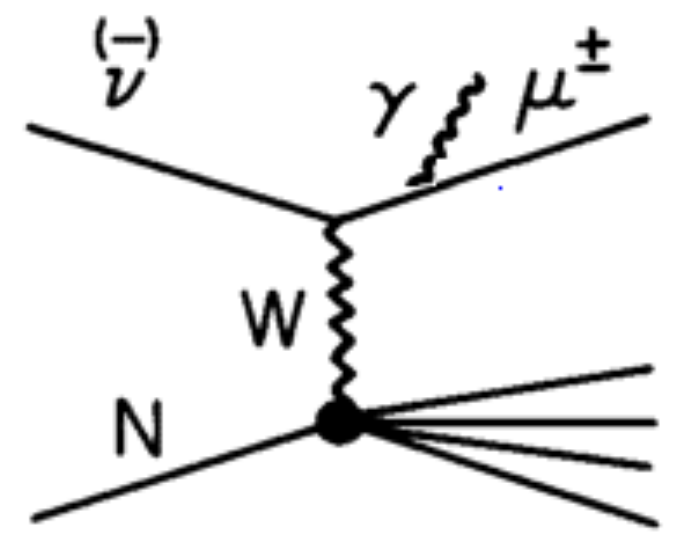}}
\caption[energy]{ The  the
bremsstrahlung diagram for radiation of a hard photon by the muon leg  $\numu$N$\rightarrow \mu\gamma$N).}
 \end{center}
     \label{fig4}  
\end{figure}

The contribution from the muon leg to the double differential cross section in
leading log approximation can then be written as

\begin{eqnarray}
  \frac{d^2\sigma}{dxdy}&=&\frac{d^2\sigma_0}{dxdy}+
  \frac{\alpha}{2\pi}\ln\frac{2ME_\nu(1-y+xy)^2}{m_\mu^2}
  \label{one}\\
  &&\times\int_0^1dz\frac{1+z^2}{1-z}\biggl[
  \frac{y\,\Theta(z-z_{\rm m})}{z(y+z-1)}
  \frac{d^2\sigma_0}{dxdy}(\tilde{x},\tilde{y})\nonumber\\
  &&\hskip2.5cm-\frac{d^2\sigma_0}{dxdy}(x,y)\biggr]\nonumber\,,
\end{eqnarray}

where $\alpha\simeq1/137$ is the fine structure constant,

\begin{eqnarray}
  \tilde{E_\mu}&=&\frac{E_\mu}{z}\nonumber\\
  \tilde{x}&=&\frac{xy}{z+y-1}\nonumber\\
  \tilde{y}&=&\frac{z+y-1}{z}\label{two}\\
  z_{\rm m}&=&1-y+xy\nonumber\,,
\end{eqnarray}
and $d^2\sigma_0/dxdy$ is the lowest order Born cross
section as shown in figure~3.  Here, $Q^2=2ME_\nu xy$,
 $M$ is the nucleon mass, $y=(E_\nu-E_\mu)/E_\nu$, and $E_\mu$ the
energy of the outgoing charged muon in the
laboratory frame. 
Eq.~(\ref{one}) is expected to be a good approximation
for $\ln(2ME_\nu/m_\mu^2)\gg\ln(1-y)$~\cite{rujula}.

\begin{table*}
\begin{center}
\begin{tabular}{|l|l|l|l|l|l||l|l|l|l|l|l|}
\hline
$E_{\mu-qe} $ & $E_{\nu}$ & $\theta_\mu$ & $Q^2$ & $P(E_\mu$$< $ & $P(E_\mu$$<$ 
& 
$E_{\mu-qe} $ & $E_{\nu}$ & $\theta_\mu$ & $Q^2$ & $P(E_\mu$$< $ & $P(E_\mu$$<$ 
\\
$GeV $ & $GeV$ & $degrees$ & $(GeV/c)^2$ & $0.05E_{\mu-qe})$& $0.5E_{\mu-qe})$
&
$GeV $ & $GeV$ & $degrees$ & $(GeV/c)^2$ & $0.05E_{\mu-qe})$&
$0.5E_{\mu-qe})$\\ 
\hline
0.117 & 0.145&	35.0&	0.01&	0.0015&	0.0006&	0.175&	0.204&	29.0&	0.01&	0.0033&
0.0016\\
0.117 &0.156&	100.0&	0.03&	0.0021&	0.0006&	0.175&	0.254&	106.5&	0.10&	0.0055&
 0.0016\\
0.117 &0.164&	140.0&	0.04&	0.0023&	0.0006&	0.175&	0.284&	140.0&	0.15&	0.0060&
 0.0016\\
0.117 &0.167&	180&	0.05&	0.0023&	0.0006&	0.175&	0.300&	180.0&	0.18&	0.0061&
 0.0016\\
\hline
0.226&     0.255 &	23.0 &	0.01 &	0.0044	& 0.0021 &
0.299&	0.328&	17.5&	0.010&	0.0054&	0.0026
\\
0.226 &	0.305 &	77.0 &	0.10 &	0.0064	& 0.0021 & 
0.299&	0.378  &57.0 &	0.100&	0.0072&	0.0027 
\\
0.226 &	0.361 &	108.5 &	0.20 &	0.0073 &	0.0021 &
0.299&	0.433&	78.2&	0.200&	0.0082&	0.0026
\\
0.226 &	0.469 &	180.0 &	0.39 &	0.0078 &	0.0020 & 0.299 &
0.909&	180.0&	1.051&	0.0084&	0.0021
\\
\hline
0.414 &	0.442&	13.0&	0.01&	0.0066&	0.0033&	0.807&	0.835&	7.0&	0.01&	0.0091&
0.0048\\
0.414&	0.492&	41.3&	0.10&	0.0081&	0.0033&	0.807&	0.885&	21.6&	0.10&	0.0100&
 0.0047\\
0.414&	0.547&	56.4&	0.20&	0.0091&	0.0033&	0.807&	0.940&	29.8&	0.20&	0.0108&
 0.0047\\
0.414&	0.714&	81.8&	0.50&	0.0103&	0.0031&	0.807&	1.105&	44.0&	0.50&	0.0122&
 0.0045\\
0.414&	0.992&	103.5&	1.00&	0.0101&	0.0027&	0.807&	1.382&	56.5&	1.00&	0.0128&
 0.0042\\
0.414&	1.216&	114.2&	1.40&	0.0096&	0.0025&	0.807&	1.604&	62.7&	1.40&	0.0128&
 0.0040\\
0.414&	1.554&	125.2&	2.00&	0.0088&	0.0022&	0.807&	1.938&	68.9&	2.00&	0.0123&
 0.0036\\

\hline				
1.703 &	1.731&	3.3&	0.01&	0.0120&	0.0067&	4.001&	4.029&	1.4&	0.01&	0.0157&
0.0093\\
1.703&	1.781&	10.4&	0.10&	0.0125&	0.0067&	4.001&	4.079&	4.5&	0.10&	0.0159&
 0.0092\\
1.703&	1.836&	14.5&	0.20&	0.0130&	0.0066&	4.001&	4.134&	6.3&	0.20&	0.0162&
 0.0092\\
1.703&	2.001&	22.1&	0.50&	0.0141&	0.0065&	4.001&	4.299&	9.8&	0.50&	0.0169&
 0.0091\\
1.703&	2.276&	29.4&	1.00&	0.0152&	0.0062&	4.001&	4.573&	13.4&	1.00&	0.0177&
 0.0089\\
1.703&	2.497&	33.3&	1.40&	0.0156&	0.0060&	4.001&	4.793&	15.5&	1.40&	0.0182&
 0.0088\\
1.703&	2.828&	37.6&	2.00&	0.0157&	0.0057&	4.001&	5.123&	18.0&	2.00&	0.0188&
 0.0085\\
\hline
10.001 & 10.029 &0.6&	0.01&	0.0200&	0.0125&	20.000&	20.029&	0.3&
0.01&	0.0236&	0.0152 \\
10.001&	10.078&	1.8&	0.10 &	0.0201&	0.0125&	20.000&	20.078&	0.9&	0.10&	0.0236&
 0.0152\\
10.001&	10.133&	2.6&	0.20&	0.0202&	0.0125&	20.000&	20.133&	1.3&	0.20&	0.0237&
 0.0152\\
10.001&	10.298&	4.0&	0.50&	0.0206&	0.0124&	20.000&	20.297&	2.0&	0.50&	0.0238&
 0.0152\\
10.001&	10.572&	5.6&	1.00&	0.0210&	0.0123&	20.000&	20.572&	2.8&	1.00&	0.0241&
 0.0151\\
10.001&	10.791&	6.5&	1.40&	0.0214&	0.0122&	20.000&	20.791&	3.3&	1.40&	0.0243&
 0.0150\\
10.001&	11.121&	7.7&	2.00&	0.0218&	0.0120&	20.000&	21.120&	4.0&	2.00&	0.0246&
 0.0149\\
\hline
\end{tabular}
\caption[radiative correction]{A few numerical examples
of the probability for a muon generated quasielastic
neutrino scattering  to radiate a high energy photon with
energy greater than 5$\%$ and 50$\%$ of the energy of
the quasielastic muon $E_{\mu-qe} =
(p_{\mu}^{2}+m_{\mu}^{2})^{1/2}$. The first two entries in 
the table, $E_{\mu-qe}$ = 0.117 GeV ($p_{\mu}$ = 0.050 $GeV/c$),
and $E_{\mu-qe}$ = 0.175 GeV ($p_{\mu}$ =0.140 $GeV/c$) are in the
LSND (decay in flight) energy region (and lower range of MiniBoone). 
The higher energy entries are  useful for experiments
that extract the axial form factor from quasielastic events. }
\label{parameters}
\end{center}
\end{table*}

The first term under the integral sign of equation (1) comes from the
bremsstrahlung diagram of figure ~4 ($\numu$N$\rightarrow$$\mu$$\gamma$N).  Muons of energy 
 $ \tilde{E_\mu}$= $E_\mu$/$z >E_\mu$ radiate photons and contribute
 to the uncorrected differential cross sections for
 outgoing muons of energy  $E_\mu$. The case of $z \rightarrow z_{\rm m}$ 
 corresponds to the case when the initial muon originates
 from quasi-elastic scattering.  The case of $z \rightarrow 1$ corresponds
 to the case where the energy of the radiated phone $E_\gamma \rightarrow 0$.
 
For MiniBoone we  make the approximation that the cross section is dominated
by quasi-elastic scattering from nucleons at rest. In that case,  the
Born cross section is a delta function at the quasielastic peak.

\begin{eqnarray}
  \frac{d^2\sigma_0}{dxdy}&=&G(x,y)\delta (x-1)
   \label{three}
  \end{eqnarray}

Where $G(Q^2$) is a combination of vector and axial form factors.
Integrating equation (1)  over the elastic delta function we
obtain the probability $dP(y)/dy$ for a muon (from
a  quasielastic  $\numu$  scattering event) with energy 
$ E_{\mu-qe}$ to radiate a hard photon and
end up in with energy $E_{\mu} =y E_{\nu}$:

\begin{eqnarray}
  \frac{dP(y)}{dy}&=&\frac{\alpha}{2\pi}\ln\frac{2ME_\nu z_{\rm m}^2}{m_\mu^2}\times
  \label{four}\\
  && \frac{1+z_{\rm m}^2}{1-z_{\rm m}}\biggl[
  \frac{y}{z_{\rm m}}\biggr]\nonumber\,,
\end{eqnarray}

We define  r=$E_\gamma /E_{\mu-qe}$.
The case of $y_{low}$ =($1- E_{\mu-qe}/E_\nu$)  is when the  energy
of the radiated photon $E_\gamma \rightarrow 0$ ($r=r_{low}=0$). 
 The
case of  $y_{high}$ =$ 1- m\mu/ E_\nu$ is when the momentum of
the final state muon is zero. The case of $r=r_{high}$  yields the highest possible
radiated photon energy  $E_\gamma \rightarrow E_{\mu-qe} -m_\mu$.
By Integrating equation (\ref{four}),  we  
 obtain the integrated probability P($E_{\gamma -min}$)
   for a muon to radiate a photon with energy $E_\gamma > E_{\gamma -min}$.

  The background for $\nue$ events originates from events
  with a high $E_{\gamma}$.  In MiniBoone, the total measured energy
of  an  electromagnetic shower for a quasi-elastic events with a
high photon energy is the sum of the energy of the photon
and the fraction of energy of the muon above Cerenkov threshold. This is given
 approximately by
  $E_{EM}$=$E_\gamma$ + Max ($E_{\mu-qe}-E_\gamma$-0.2 GeV, 0).

We now give a few numerical examples.   For the case of 
$E_\nu$=  442 MeV,  $E_{\mu-qe}$ = 414 MeV,  and $\theta_\mu$ =  13 degrees
 ($Q^2$=0.01 $(GeV/c)^2$), the probability for radiating a hard photon with
 an energy which is higher than a typical experimental muon momentum resolution (e.g. $5\%$)
 is 0.66$\%$. The probability for radiating a photon with $E_\gamma > 207~MeV$  (i.e.
 50$\%$ of its energy) 
 is 0.33$\%$.  This rate an order of magnitude larger than the 0.03$\%$ rate of excess
 $\nue$ candidates in MiniBoone.  Additional numerical examples  at other
 representative energies are given in Table (I) including both  lower energy
 (e.g. 226 MeV corresponding to the lower energy range at  MiniBoone and the upper range of
 LSND), as well as  higher  energy muon.

   In the MiniBoone experiment the nucleons are bound in carbon.
For the case of quasi-elastic  $\nu$ scattering,  a 
   a nucleon of mass $M_1$ ( bound in $C^{12}$) is scattered to a 
 final state nucleon with mass $M_{2}$. Here, 
 $Q^2 = \vec{q}^{~2}-(E_\nu-E_{l})^{2}= -m_l^{2} + 2 E_\nu(E_{l} - p_l \cos\theta_\mu)$,
where $E_{l}$  is the final state lepton energy (muon or electron).   MiniBoone
  calculates the  reconstructed energy  (shown in figure~1) as follows.
\begin{eqnarray}
     E_{\nu}^ {rec}= \frac{2(M_1 - E_B)E_{l} - (E_B^2 - 2M_1 E_B + m_{l}^{2} +
             \Delta M^{2})}
             {2\:[(M_1 - E_B) - E_{l} + p_{l}
	     \cos\theta_{l}]}
    \label{five}\
\end{eqnarray}
MiniBoone uses an average removal energy
$E_B= ~37 MeV$. Here,  $m_{l}$ is final state lepton mass,
$p_{l}$ is the final state lepton momentum, and  $\Delta M^{2} = M_1^2 -
M_2^2$.   For $\nue$ candidates, the final state lepton mass
  is assumed to be the  electron mass (which can be neglected). 
  When calculating the reconstructed
  energy for the radiative muon background to the  $\nue$  candidates, the final state
  energy and momentum of
  the lepton is $E_{EM}$
  Therefore, the reconstructed neutrino energy (shown in figures 1 and 2)  is higher than  $E_{EM}$ for
  electromagnetic showers at larger angles. 
  
 Although radiative muon events are a large source of background in 
 $\numu$$\rightarrow$$\nue$
neutrino oscillations appearance experiments, radiative corrections have
only a small effect on the overall quasielastic differential cross section
at low energies.  However, as the precision of the next generation neutrino experiments 
improves, radiative corrections  should be accounted for. 
  It is usual to define $\delta$ = P ($E_{\gamma-min} = \Delta E_\mu$) as the radiative
correction to quasielastic scattering. Here  $\Delta (E_\mu)$ = $\sigma \times E_{\mu-qe}$, where
$\sigma$  is  larger than the experimental error on the measurement
of the energy of the final state muon,. 
   The Born quasielastic cross section
 $ \frac{d^2\sigma_{0-qe}}{dy}$
is related to the measured quasielastic 
cross section $ \frac{d^2\sigma_{quasi}}{dy}$via the following expression.
\begin{eqnarray}
  \frac{d^2\sigma_{quasi}}{dy}&=&\frac{d^2\sigma_{0-qe}}{dy}\biggl[{1 - \delta (\Delta E_\mu)} \biggr]\
 \label{six}
\end{eqnarray}
  The above expression can be used to correct the value
  of the axial vector mass $M_A$ extracted from quasielastic
  neutrino scattering experiments for radiative effects. In general,
  correcting for radiative effects will increase extracted value of $M_A$
  (since correcting for radiative effects yields is a larger correction
  at larger $Q^2$).  The values of the radiative corrections for an experiment
  with a typical muon energy resolution  $\Delta= 0.05 \times E_{\mu-qe}$ are also given in Table (I). 
  The   updated world
average value\cite{bbba2007}  of  the axial vector mass $M_A$ extracted  from $\numu$
  quasielastic scattering bubble chamber experiments with deuterium is  $M_{A}^{deuterium}$ =$1.014 \pm 0.026$  $GeV/c^2$.  For a typical $\numu$ energy of 2 GeV, the corrections
  for radiative effect given in Table (I) implies that this average value
  should be increased by   $\delta M_A^{\numu}  \approx 0.002$.   For bubble chamber
  experiments with antineutrinos, the correction depends on the range
  of $Q^2$ of the analysis. 

In summary, we show that the background from
$\numu$N$\rightarrow$$\mu$$\gamma$N can account
for the excess $\nue$ candidates in the  MiniBoone
and LSND (decay in flight) experiments.
 Hard photons radiated by the muon leg in charged 
current events  are a significant source of
background that should be considered by $\numu$$\rightarrow$$\nue$
neutrino oscillations appearance experiments (e.g. LSND, MiniBoone, SuperK, MINOS,  T2K and NOVA).

\end{document}